\documentclass[twocolumn,showpacs,preprintnumbers,nofootinbib]{revtex4-1}
\usepackage[utf8]{inputenc}
\usepackage{amsmath}
\usepackage{amsthm}
\usepackage{amssymb}
\usepackage{enumitem}   
\usepackage[english]{babel}
\usepackage{hyperref}
\usepackage{mathtools}
\usepackage{bbold,xcolor}
\usepackage[normalem]{ulem}

\newcommand{\SO}{\mathrm{SO}}
\newcommand{\SU}{\mathrm{SU}}
\newcommand{\UH}{\mathrm{U}(1)_{\mathrm{H}}}
\newcommand{\U}{\mathrm{U}}

\begin{document}

\author{Tommi Alanne}
\email{tommi.alanne@mpi-hd.mpg.de}
\author{Simone Blasi}
\email{simone.blasi@mpi-hd.mpg.de}
\author{Florian Goertz}
\email{florian.goertz@mpi-hd.mpg.de}
\affiliation{Max-Planck-Institut f{\"u}r Kernphysik, Saupfercheckweg 1, 69117 Heidelberg, Germany}

\title{A Common Source for Scalars:\\
Axiflavon-Higgs Unification}

\begin{abstract}
We propose a unified model of scalar particles that addresses the flavour hierarchies,
solves the strong CP problem, delivers a dark matter candidate, 
and provides the trigger for electroweak symmetry breaking.
Besides furnishing a unification of the recently proposed axiflavon with a Goldstone-Higgs sector,
the scenario can also be seen as adding a model of flavour (and strong CP conservation along 
with axion dark matter) to elementary Goldstone-Higgs setups.
In particular, we derive bounds on the axion decay constant from the need to generate a
SM-like Higgs potential at low energies, which we confront with constraints from flavour physics 
and cosmology.
In the minimal implementation, we find that the axion decay constant is restricted to a 
thin stripe of $f_a \approx (10^{11}-10^{12}) \, \text{GeV}$,
while adding right-handed neutrinos allows to realize a \emph{heavy-axion} model at lower energies, down to $f_a \sim 10$ TeV.

\end{abstract}

\maketitle

\section{Introduction}

Although the Standard Model (SM) of particle physics
provides an excellent description of nature around the
weak scale, it has several shortcomings that lead 
us to the conclusion that it is rather an effective 
low-energy parametrization of a more fundamental 
theory of nature.

Amongst the most pressing issues are the missing
candidate to generate the dark matter 
populating our universe, and the failure to explain
large hierarchies present in the fermion masses 
and mixings. Beyond that, the apparent conservation 
of CP symmetry in strong interactions is in 
tension with in principle unsuppressed sources
of CP violation in the QCD Lagrangian. 
Finally, although the SM provides a successful 
parameterization of electroweak symmetry
breaking (EWSB) via the Higgs mechanism, 
the origin of the Higgs 
potential is unknown.

The flavour hierarchies in the SM can be addressed
via the Froggatt--Nielsen (FN) mechanism \cite{Froggatt:1978nt}, i.e. by
chirally charging the SM fermions under a $\UH$
flavour symmetry controlling their masses and mixings.
For the FN mechanism to work, the field content of the
SM has to be augmented (at least) by a complex 
scalar field, $\Phi$, spontaneously breaking the 
$\UH$, and vector-like fermions, $\xi_j$, dubbed the FN
messengers, 
which connect the different SM-fermion chiralities,
bridging their $\UH$ charge difference via a chain
of $\Phi$ insertions. Taking the FN messengers to
reside much above the electroweak (EW) scale, they can
be integrated out in the IR description.
Then, once the $\UH$ is spontaneously broken by 
$\langle \Phi \rangle \neq 0$, the SM Yukawa couplings are
effectively reproduced starting with $\mathcal{O}(1)$
couplings in the UV theory.

As shown in a previous work~\cite{Calibbi:2016hwq} (see
also Ref.~\cite{Ema:2016ops}), the angular component of 
$\Phi$, which plays no vital role in the original FN
mechanism,
can be
identified with the QCD axion~\cite{Peccei:1977hh,Wilczek:1977pj,
Weinberg:1977ma,Kim:1979if,Shifman:1979if,Dine:1981rt,Zhitnitsky:1980tq},
thereby addressing two
more issues of the SM, namely the strong CP problem and
the DM puzzle~\cite{Preskill:1982cy,Abbott:1982af,Dine:1982ah}, 
in a unified scenario. In particular, as
the axion couplings are now dictated by the flavour
structure, the predictivity of the model is increased.

In this paper, we take
a further step to unify the
scalar degrees of freedom in the theory by combining
the axiflavon field, $\Phi$, and the Higgs boson such
that the flavour, strong CP, and DM problems remain
solved, while successfully triggering EWSB.
We thus embed the Higgs, the axion, and the
flavon in a single 
multiplet, $\Sigma$,
transforming under the enlarged symmetry group 
$\mathcal{G} \supset \UH \times G_\text{EW}$,
with $G_\text{EW} \equiv \SU (2)_{\mathrm{L}} \times \U (1)_Y$.
Within the FN setup, the flavon mass
and the Higgs mass are hierarchically different. 
In the unified picture, this very fact suggests that
both the axion and the Higgs boson components should
correspond to pseudo-Nambu--Goldstone bosons (pNGBs),
providing an example of axion-Higgs unification
\cite{Redi:2012ad} and allowing for dynamical EWSB via
the Coleman-Weinberg potential for the emerging
Goldstone Higgs. Interestingly, the vanishing of
the quartic Higgs coupling in the SM around $10^{10}$\,GeV, 
just about at the natural scale for axiflavon dark matter, 
might hint both to Goldstone nature of the Higgs
and to a connection between these two scalar sectors. 

From a different perspective, the proposed model
can be seen as adding a flavour story to the
recently proposed elementary-Goldstone-Higgs 
scenario~\cite{Alanne:2014kea,Alanne:2015fqh}. 
Including the
axiflavon 
can
address fermion masses and mixings in these models, while
providing a solution to the strong CP problem.
This is a compelling renormalizable alternative
to partial
compositeness \cite{Kaplan:1991dc} generating flavour hierarchies
in composite-Higgs models.
Furthermore, in this way the flavour structure 
can be 
achieved without the
need of adding a new disconnected scalar or a new
symmetry-breaking mechanism. 
We stress that the 
fine-tuning problem
affecting the EW scale in this setup is not worse
than the one in the usual FN mechanism, which already
involves the tuning of the Higgs portal
coupling $\lambda |H|^2 |\Phi|^2$. 
At the end of the paper, we also comment on
the option to realize the model around the
TeV scale.

Before working out the setup and its predictions in
detail, we summarize the main model building 
steps. We formulate the theory at the scale $f$ as
a linear sigma model for the field $\Sigma$. 
Yukawa interactions between $\Sigma$ and the FN
messengers are introduced as a microscopic realization
of the FN mechanism, and similarly for the SM fields. 
The global symmetry breaking pattern is
$ \mathcal{G} \xrightarrow{\text{\tiny $\langle \Sigma \rangle$}} \mathcal{H} \supset G_\text{EW}$, 
with the axion and the Higgs residing in the 
$\mathcal{G}/\mathcal{H}$ coset.
Notice that $\mathcal{H}$ may or may not contain
custodial protection, as the scale $f$ is by
construction much larger than the Higgs vacuum
expectation value, so that the custodial-breaking effects are strongly suppressed.
 The simplest choice for
$\mathcal{G}$ is to keep the flavour symmetry as
an abelian factor, $\mathcal{G} = \mathcal{G^\prime} \times \UH$.
The minimal choice for 
$\mathcal{G^\prime}$ is based on $\SU(3)$ and the 
next-to-minimal on $\SO(5)$ \cite{Agashe:2004rs}. Both result in a
similar structure, and in the following we
focus only on the latter. 
Finally, we assume that the explicit
breaking of $\mathcal{G}$ originates from
the SM sector only, namely from the QCD anomaly, EW gauging,
and via the SM fermions coming as $\mathcal{G}$
spurions.
Conversely,
the FN messengers always enter as full
representations.

This article is organized as follows.
In Section \ref{sec:setup}, we detail the
setup and main features of the model, including
the structure of the linear sigma model and the
generation of mass hierarchies via the
FN mechanism. Section
\ref{sec:Pot} contains the calculation and
analysis of the Higgs potential, which leads
to a prediction for the axion decay constant.
Here, we also discuss the impact of including 
right-handed neutrinos into the setup and
potential constraints from flavour physics and cosmology. Finally,
Section \ref{sec:con} contains our conclusions.
In two appendices we discuss the assumptions
regarding the mass spectrum of the FN messengers
as well as the contributions from light fermions 
to the Higgs potential.

\section{Model Setup}

\label{sec:setup}

In the following, we present the explicit model setup. 
The symmetry-breaking pattern that leads to the unified realization of 
the Higgs doublet and the axion as pNGBs reads
\begin{equation}\label{eq:so5pattern}
 \left[ \SO(5) \times \UH \right] \times \U(1)_X \rightarrow \SO(4)
 \times \U(1)_X\,,
\end{equation}
where the $\U(1)_X$ factor is introduced to reproduce 
the fermion hypercharges.
The pattern of Eq.~\eqref{eq:so5pattern} is obtained within 
a linear $\sigma$-model for the field $\Sigma$ living in the fundamental representation,
$\mathbf{5}$, of $\SO(5)$ and having $\UH$ flavour charge $H_\Sigma = 1$, with the potential
\begin{equation}\label{eq:potential}
 V(\Sigma, \Sigma^*) = 
 \lambda_1 \left(\Sigma^\dagger \Sigma \right)^2 - 
 \lambda_2 \, \Sigma^T \Sigma \, \Sigma^\dagger \Sigma^*  - \mu^2 \Sigma^\dagger \Sigma\,.
\end{equation}
The EW gauge group is embedded in $\SO(5)$ by defining the usual 
$\SU(2)_{\mathrm{L}} \times \SU(2)_{\mathrm{R}} \cong
\SO(4)$ generators:
\begin{equation}
 T_{ij}^{a \, \mathrm{L,R}} = -\frac{i}{2} \left[ \frac{1}{2} \epsilon^{abc}
 (\delta^b_i \delta^c_j - \delta^c_i \delta^b_j) \pm
 (\delta_i^a \delta_j^4 - \delta_j^a \delta_i^4) \right].
\end{equation}
The $\SO(4)$-preserving minimum is then given by
$\langle \Sigma \rangle = (0,0,0,0,f/\sqrt{2})$, with
$ \mu^2 = (\lambda_1 - \lambda_2) f^2$, and after the breaking, Eq.~\eqref{eq:so5pattern},
the scalar
sector can be parametrized as
\begin{equation}
\Sigma = e^{i (\sqrt2 h_{\hat a} \hat T^{\hat a} + a) /f} \begin{pmatrix} \widetilde{H} \\ (f + \sigma)/\sqrt 2 \end{pmatrix}\,,
\end{equation}
where the broken generators, $\hat T^{\hat a}$, are given by 
\begin{equation}
    \label{eq:}
    \hat T_{ij}^{\hat a} = - \frac{i}{\sqrt 2} \left[ \delta_i^{\hat a}  \delta_j^5 - \delta_j^{\hat a}  \delta_i^5\right].
\end{equation}
As physical states, one finds a heavy Higgs doublet, $\widetilde{H}$, with mass 
$m^2_{\tilde H} = 2 \lambda_2 f^2$, and a heavy flavon,
$\sigma$, with mass $m^2_\sigma = 2 (\lambda_1 - \lambda_2) f^2$, 
while the SM-like Higgs doublet, $h_{\hat a}$, and the axion, $a$, are instead pNGBs.
The potential is bounded from below if 
$\lambda_1 > \lambda_2 > 0$.

The FN messengers are denoted by $\xi_j$, where the 
subscript refers to the $\UH$ charge, $H_{\xi_j}=j$. 
Each of them transforms in the spinorial representation, $\mathbf{4}$, 
of $\SO(5)$ and is vectorial under $\SO(5) \times \UH$.

Similarly, the SM fermions, $q_{\mathrm{L}}^i$, $u_{\mathrm{R}}^i$, and $d_{\mathrm{R}}^i$
(with $i=1,2,3$ a flavour index),
are introduced as spurions, $\Psi_f^i$, in the spinorial
representation:
\begin{equation}\label{eq:spurion}
 \Psi^i_{q_{\mathrm{L}}} = \Delta_{\mathrm{L}}^T q^i_{\mathrm{L}}, 
 \quad \Psi^i_{u_{\mathrm{R}}} = \Delta_u^T u^i_{\mathrm{R}},
 \quad \Psi^i_{d_{\mathrm{R}}} = \Delta_d^T d_{\mathrm{R}}^i,                  
\end{equation}
where 
\begin{equation}
\begin{split}
    \Delta_{\mathrm{L}} &= \left( \begin{array}{cccc}
                    1 & 0 & 0 & 0\\
                    0 & 1 & 0 & 0 \\
                   \end{array} \right),\\
    \Delta_{u} &= \left( 
                                          0, 0, 1, 0
                                        \right), \quad
    \Delta_{d} = \left( 
                                          0,0,0,1
                                         \right).
    \end{split}
\end{equation}
The $\UH$ charge of each $\Psi^i_f$ is chosen such that
the correct pattern of masses and mixings is reproduced.
The larger the charge difference between the left- and 
right-handed components of a given fermion, the more suppressed
is the resulting mass term. 
Notice that the $\Psi$-fields have to be considered only as $\SO(5)$ spurions, 
while the $\UH$ needs to be exact at the Lagrangian level.
For both the SM fermions and the FN messengers, 
the $\U(1)_X$ charge is chosen to match the correct
hypercharge, $Y = T_3 + X$, and will be omitted in the following.

The Lagrangian of the system includes renormalizable
operators made out of $\Psi_f^i$, $\xi_j$, and 
$\Sigma$ allowed by symmetries:
\begin{equation}
\begin{split}
 - \mathcal{L} \, & = \, \sum_j \left(a_j \bar{\xi}_{j+1} \, \Gamma^\alpha 
 \, \Sigma_\alpha \, \xi_j + \text{h.c.} \right)
 + m_j \, \bar{\xi}_j \, \xi_j,\\
 &+  \sum_{i,f} 
 z_i^f \, \bar{\Psi}^i_f \, \Gamma^\alpha 
 \, \Sigma_\alpha \, \xi_{j} + \tilde z_i^f \, \bar{\xi}_{j + 2}
 \, \Gamma^\alpha \, \Sigma_\alpha \, \Psi^i_f 
 +
 \text{h.c.} \\
 & + x \, \bar{\Psi}^3_{q_L} \, \Gamma^\alpha 
 \, \Sigma_\alpha \, \Psi^3_{u_R} + \text{h.c.}\,,
 \end{split}
\end{equation}
where $\Gamma^a$ are the matrices defining the
spinorial representation.
In the Lagrangian above, the first line contains the
interactions of the FN messengers with the $\Sigma$-field and their (vector-like) mass terms, 
while the second line consists of Yukawa
couplings involving the SM fermions and the 
FN messengers,
where $f =q_{\mathrm{L}}, u_{\mathrm{R}}, d_{\mathrm{R}}$, and $j\equiv j(f,i) = H_{f^i}-1$ is such that the terms are $\UH$ invariant.
 Notice that the use of the
spinorial representation is particularly suitable 
for the purpose of building the FN chain: both 
$\xi$-fields appear symmetrically in the Yukawa
coupling and thus only a single species of heavy fermions is needed. 
Finally, the last line accounts for the fact that 
the top mass features no suppression, and thus a direct coupling of $t_{\mathrm{L}}$ 
and $t_{\mathrm{R}}$ via the
$\Sigma$-field must be allowed.

Before presenting the computation of the Higgs potential, let us discuss
an example to show
how the FN mechanism is explicitly realized in our setup.

\subsection{Mass hierarchies from broken \text{\boldmath $\UH$}}

Consider two chiral fermions,
$\Psi_{q_{\mathrm{L}}}^3$
and $\Psi_{u_{\mathrm{R}}}^2$ for concreteness,
as given in Eq.~\eqref{eq:spurion},
and two FN messengers, $\xi_{2,3}$, with the 
mass-mixing Lagrangian
\begin{equation}\label{eq:topcharm}
    \begin{split}
	- \mathcal{L}  \supset & \,  
	    z \, \bar{\Psi}_{u_{\mathrm{R}}}^2 
	    \Sigma^\prime \xi_3 + \tilde z \, 
	\bar{\xi}_{2} \Sigma^\prime \Psi_{q_{\mathrm{L}}}^3
	+ a_{2} \, \bar{\xi}_3 \Sigma^\prime \xi_{2} + \text{h.c.} 
	\\
	&+ m \left( \bar{\xi}_{2} \xi_{2} + \bar{\xi}_3 \xi_3 \right),
    \end{split}
\end{equation}
where we have defined $\Sigma^\prime \equiv \Gamma^\alpha \, \Sigma_\alpha$.
The Lagrangian above corresponds to the flavour charges 
$H_{q_\mathrm{L}^3} = 1$ and 
$H_{u_\mathrm{R}^2} = 4$ and, as we shall see,
reproduces the term in Eq.~\eqref{eq:Lupquarks} for the
$t_{\rm L}$-$c_{\rm R}$ mixing.
All dimensionless couplings are assumed to be
$z \sim \tilde z \sim a_{2} \sim \mathcal{O}(1)$.

By integrating 
out $\xi_2$ and $\xi_3$ at the tree level, 
one finds the effective Lagrangian, $\mathcal{L}_{\text{eff}}$, which, at
the leading order in $1/m$, reads\footnote{
\label{fn:Gamma} Note that
for the apparently more minimal chain between two light fermions that differ only by 
$\Delta H=2$ units of flavour charge, the corresponding effective Lagrangian 
$\sim  \bar{\Psi}_{q_{\mathrm{L}}}^3
 \Sigma^\prime \Sigma^\prime
 \Psi_{u_{\rm R}}^2  = \bar{q}_{\mathrm{L}}^3 \Delta_{\mathrm{L}}
 \Sigma^\prime \Sigma^\prime
 \Delta_u^T \, u_{\mathrm{R}}^2$
vanishes due to $\Sigma_\alpha \Sigma_\beta \Gamma^\alpha \Gamma^\beta =
\Sigma^\alpha \Sigma_\alpha \mathbb{1}$ and $\Delta_{\mathrm{L}} \Delta_u^T = 0$.}
\begin{equation}\label{eq:Leff2}
 - \mathcal{L_{\text{eff}}} = z \, \tilde z \, a_{2} \, \frac{1}{m^2}
 \bar{\Psi}_{u_{\mathrm{R}}}^2
 \Sigma^\prime \Sigma^\prime \Sigma^\prime 
 \Psi_{q_\mathrm{L}}^3 + \text{h.c.} 
\end{equation}
Below the symmetry-breaking scale and after integrating 
out the flavon and the second Higgs doublet, 
the $\Sigma$-field can be written 
by the Goldstone parametrization in the unitary gauge as
\begin{equation}\label{eq:sigmagoldstone}
 \Sigma = \frac{f}{\sqrt{2}} \, e^{i a/f} \left(
                          0, 0, \text{sin} \, h/f
                          , 0, \text{cos}\, h/f
                         \right)^T,
\end{equation}
where $h$ represents the Higgs field, 
and $a$ is the axion.
Using Eq.~\eqref{eq:sigmagoldstone}, we can
single out the contribution to the mass matrix in
 Eq.~\eqref{eq:Leff2}:
\begin{equation}\label{eq:Leffmass}
 - \mathcal{L_{\text{eff}}} \supset m_{32} \bar{c}_{\rm R} t_{\rm L}  
 + \text{h.c.} \,,
\end{equation}
\begin{equation}
 m_{32} = \frac{1}{\sqrt{2}} z \, \tilde z \, a_{2} \frac{f^2}{2 m^2} 
f \, \text{sin}(\langle h \rangle /f) + \dots \, ,
\end{equation}
where $f \, \sin (\langle h\rangle/f)\equiv v$  is to be identified with the EW scale
and the dots stand for higher orders in $f^2/2 m^2$.
Defining 
$\delta_{ij} \equiv H_{q_\mathrm{L}^i} - H_{q_\mathrm{R}^j}$,
we see that the suppression with respect to the top mass is
$ m_{ij}/m_t \sim 
 \left( f/\sqrt{2} m \right)^{|\delta_{ij}| - 1}$
with $\delta_{32} = 3$ in the present case.
The addtional $-1$ in the exponent, 
which is not present in the usual FN setup, 
compensates for the Higgs carrying one unit of 
flavour charge, since it is unified in the $\Sigma$-field.

One can show that this result holds in general
for odd $|\delta_{ij}|$, while for even $|\delta_{ij}|$
the corresponding term vanishes due to the properties
of the $\Gamma$ matrices; see footnote \ref{fn:Gamma}.
We thus conclude that a general entry in the fermion mass matrix, $m_{ij}$,
corresponding to a charge difference of $|\delta_{ij}|$
is suppressed with respect to the top mass by
\begin{equation}\label{eq:suppr}
  \frac{m_{ij}}{m_t} \sim
       \left( \frac{f^2}{2 m^2} \right)^{\frac{|\delta_{ij}| - 1}{2}}\,,
       \quad |\delta_{ij}| \, \text{odd}\,.
\end{equation}
Eq.~\eqref{eq:suppr} shows that
$\left(f^2 / 2 m^2\right) \equiv \epsilon$ is the smallest building block
we can use to reproduce the flavour hierarchies, and therefore
we identify $\epsilon = \text{sin}\,\theta_{\mathrm{C}} \simeq 0.23$ with
$\theta_{\mathrm{C}}$ being the Cabibbo angle.
As a final check,
for the $t_{\rm L}$-$c_{\rm R}$ mixing with $\delta_{32}=3$,
we see that Eq.~\eqref{eq:suppr} gives the 
correct entry $n_{32} = 1$ in Eq.~\eqref{eq:nij}.

\section{Higgs Potential and Constraints on the Axion Decay Constant}
\label{sec:Pot}

In this section, we compute the Higgs potential
generated by the interaction with the top quark and the FN messengers which directly couple to it. 
A charge assignment that is compatible
with the top mass must satisfy $|\delta_{33}| \equiv |H_{q_{\rm L}^3} - H_{u_{\rm R}^3}| = 1$,
and we take $H_{q_{\rm L}^3} = 1$, and $H_{u_{\rm R}^3} = 2$.
This corresponds to
\begin{equation}\label{eq:xicsi}
\begin{split}
  - \mathcal{L}  \supset &
  \left(a_0 e^{i \alpha}
 \bar{\xi}_1 \Sigma^\prime \xi_0 + \text{h.c.}\right)
 \, + m (\bar{\xi_0} \xi_{0} + \bar{\xi}_1 \xi_1) \\[1mm]
  & + \left( x  e^{i \zeta} \bar{\Psi}^3_{q_{\rm L}} \Sigma^\prime \Psi^3_{u_{\rm R}} 
 +  z_{\rm L} e^{i \zeta_{ \rm L}} \bar{\Psi}^3_{q_{ \rm L}} \Sigma^\prime \xi_0 \right.\\
 &\left.\quad+  z_{\rm R} e^{i \zeta_{\rm R}} \bar{\Psi}^3_{u_{\rm R}} \Sigma^\prime \xi_1
 + \text{h.c.}\right)\,,
 \end{split}
\end{equation}
where $z_{\mathrm{L,R}} \equiv z_3^{q_{\mathrm{L}},u_{\mathrm{R}}}$ 
and potential phases have 
been pulled out into $\zeta_{\mathrm{L,R}}$.
Note that we are here assuming mass degeneracy 
for the messenger fields, $m_0 = m_1 \equiv m$, 
and we are setting to zero interactions 
involving other FN messengers
which have to be there to reproduce the 
correct mixings (e.g., the 
$\bar{\xi}_2 \Sigma^\prime \Psi_{q_{\rm L}}^3$ 
term in
Eq.~\eqref{eq:topcharm})
but do not alter the conclusion of this section. 
We comment on breaking the mass degeneracy in Appendix \ref{app:m}, while
the contribution to the Higgs potential
from sectors involving lighter fermions 
is discussed in Appendix \ref{app:sub}.

Let us first compute the top mass according to
Eq.~\eqref{eq:xicsi} in the background
of $h$.\footnote{We use the same 
symbol for the classical background as for the Higgs field earlier
in Eq.~\eqref{eq:sigmagoldstone} to simplify the notation.}
This can be done by solving the characteristic
polynomial for the fermion mass matrix, 
or alternatively by integrating out the heavy fields
at the tree level.
The expression for $m_t$ is obtained as an 
expansion in $f^2/2 m^2$ and $h/f$:
\begin{equation}\begin{split}\label{eq:topmass}
 m_t^2(h)= &  \left(\alpha_0 - \alpha_2 \frac{f^2}{2 m^2} 
 - \alpha_4 \frac{f^4}{4 m^4} \right)
 \frac{f^2}{2} \text{sin}^2(h/f) \\
& -\beta_4 \frac{f^4}{4 m^4} \frac{f^2}{2} \text{sin}^4(h/f) 
+  \text{h.o.},
\end{split}
\end{equation}
where h.o. stands for higher-order contributions 
in $f^2/2m^2$ and $(h/f)^n$ terms with $n \geq 6$, and
\begin{equation}
 \begin{split}
\alpha_0 = &\, x^2, \\  
 \alpha_2 = & \, x^2\,\left(z_{\mathrm{L}}^2 + z_{\mathrm{R}}^2\right) - 2 \, 
    x \, z_{\mathrm{L}} \, z_{\mathrm{R}} \, a_0 \, \cos\,\Omega , \\
 \alpha_4 = & - x^2 \, \left(z_{\mathrm{L}}^4 + z_{\mathrm{L}}^2 z_{\mathrm{R}}^2 
     + z_{\mathrm{R}}^4\right) \\
&+ a_0^2 \left(3 x^2 z_{\mathrm{L}}^2 + z_{\mathrm{R}}^2 (3 x^2 - z_{\mathrm{L}}^2)\right) \\ 
& - 2 a_0 x z_{\mathrm{L}} z_{\mathrm{R}} \left(a_0^2 - z_{\mathrm{L}}^2 - 
    z_{\mathrm{R}}^2\right) \cos\,\Omega, \\
\beta_4 = &  \, x^4 \left(z_{\mathrm{L}}^2 + z_{\mathrm{R}}^2\right) 
    - 6 x^3\, z_{\mathrm{L}} \, z_{\mathrm{R}} \,a_0 \,\cos\,\Omega,
 \end{split}
\end{equation}
with $\Omega \equiv \alpha - \zeta_{\mathrm{L}} + \zeta - \zeta_{\mathrm{R}}$.
The expression of Eq.~\eqref{eq:topmass} needs to coincide
with the SM result, $m_t^2 = \frac{1}{2} y_t^2 h^2$
implying $y_t \simeq x$ at the leading order.

We compute the Higgs potential
by matching the SM effective
potential 
renormalized at the scale $m$ (where the new physics kicks in) 
with the one in the axiflavon-Higgs scenario.
We work out  the one-loop matching explicitly in the following.
The SM effective potential up to one-loop level, keeping only 
the top contribution, reads
\begin{equation}
    \begin{split}
	V_\text{SM}^{(1)}= &\frac{1}{4} \lambda(m) h^4 - \frac{1}{2} \mu^2(m) h^2\\ 
	&- \frac{N_c}{16\pi^2} m_t^4(h)\left( \log \frac{m^2_t(h)}{m^2}-\frac{3}{2}\right),
    \end{split}
\end{equation}
where $N_c=3$, and all the couplings are evaluated at the scale $m$,
including $y_t(m)$ within $m_t(h)$.

In the axiflavon-Higgs picture, the Higgs potential
arises at the loop level, and
it is given in terms of the field-dependent masses
of the physical eigenstates, namely the SM particles 
and the FN messengers. Considering only the top sector
in Eq.~\eqref{eq:xicsi} yields
\begin{equation}
\label{eq:CW2}
    \begin{split}
	 V_\text{AFH}^{(1)} =& - \frac{N_c}{16\pi^2} \left\{m_t^4(h)
	    \left( \log \frac{m^2_t(h)}{m^2}-\frac{3}{2}\right)\right.\\
	 &\qquad\ \   \left.+ \sum_j m_{\xi_j}^4(h)
	 \left( \log 
	 \frac{m^2_{\xi_j}(h)}{m^2}-\frac{3}{2}\right) \right\}.
    \end{split}
\end{equation}
Then, by requiring
\begin{equation}
\label{eq:match1}
 V_\text{SM}^{(1)} = V_\text{AFH}^{(1)},
\end{equation}
we obtain
\begin{equation}
    \label{eq:match2}
    \begin{split}
	 V_4 \equiv &\frac{1}{4} \lambda(m) h^4 - \frac{1}{2} \mu^2(m)h^2 \\
	 =& -\frac{N_c}{16 \pi^2} \sum_j m_{\xi_j}^4(h)
	 \left( \log 
	 \frac{m^2_{\xi_j}(h)}{m^2}-\frac{3}{2}\right)\,.
    \end{split}
\end{equation}
A few comments are in order: the contribution from the top
eigenstate appears on both side of Eq.~\eqref{eq:match1}, and therefore it cancels.
Such cancelation takes care of the large logarithm
arising from the hierarchy between the EW 
and the flavon scale.
We expect this behaviour to persist beyond the one-loop level,
given that the pure SM contribution always
appears on both sides, so that
Eq.~\eqref{eq:match2} gives the 
leading-order matching condition.

To compute the RHS of Eq.~\eqref{eq:match2}, 
we parametrize the field-dependent FN masses as
\begin{equation}\label{eq:FNeigenvalues}
 m_{\xi_j}^2(h) = m^2 + f_j(h),
\end{equation}
where $f_j(h) \sim f \times m$.
By expanding the logarithm, we see that
\begin{equation}
    \label{eq:logexpansion}
    \begin{split}
	 V_4= - \frac{N_c}{16 \pi^2} \sum_j &\left[ -2 f_j(h) m^2 + 
	 \frac{f_j^3(h)}{3 m^2} - \frac{f_j^4(h)}{12 m^4} \right.\\
	 &\left.\ + \frac{f_j^5(h)}{30 m^6} 
	 -\frac{f_j^6(h)}{60 m^8} + \text{h.o.}
	 \right],
    \end{split}
\end{equation}
where we dropped a constant term.
The computation of $F_n = \sum_j f_j^n(h)$, $n = 1,\dots,6$,
can be done recursively as
\begin{equation}\label{eq:recurs}
\begin{split}
 &F_1 =  \sum_j f_j(h) =  \text{Tr} \left[m^\dagger(h) m(h)\right] - m_t^2(h), \\
 &F_2 =  \sum_j f_j^2(h) = \text{Tr} \left[ \left(m^\dagger(h) m(h)\right)^2 \right]\\
&\hspace{2.7cm} - 2 F_1 m^2 - m_t^4(h), 
\end{split}
 \end{equation}
and so on.

By direct inspection, the terms $F_5$ and $F_6$ turn out to be 
higher order.
We eventually find for the RHS of Eq.~\eqref{eq:match2}:
\begin{equation}
    \begin{split}
	 V_4 =& \frac{N_c}{16 \pi^2} f^2 
	 \left( \gamma_0 + \gamma_2 \frac{f^2}{2 m^2} \right) \frac{f^2}{2} \text{sin}^2(h/f)\\
	 &+\frac{N_c}{16 \pi^2} f^2 \delta_2 \frac{f^2}{ 2 m^2} \frac{f^2}{2} \text{sin}^4(h/f),
    \end{split}
 \end{equation}
where
\begin{equation}
    \begin{split}
	\gamma_0 = &  \alpha_2, \\
	\gamma_2 = & \frac{1}{3} \left(-3 x^2\left(z_{\mathrm{L}}^4+z_{\mathrm{L}}^2z_{\mathrm{R}}^2
	    +z_{\mathrm{R}}^4\right)\right. \\ 
	&\quad+ a_0^2\left(3 x^2 z_{\mathrm{L}}^2
	    + z_{\mathrm{R}}^2\left(3 x^2-z_{\mathrm{L}}^2\right)\right)\\
	&\left.\quad+2a_0^3 z_{\mathrm{L}} x z_{\mathrm{R}} \, \cos\,\Omega \right), \\
	\delta_2 = & \beta_4.
    \end{split}
\end{equation}
The matching then requires
\begin{equation}\label{eq:mu2}
 \mu^2(m) = - \frac{N_c}{16 \pi^2} f^2 \left( \gamma_0 + 
 \gamma_2 \frac{f^2}{2 m^2} \right)\,,
\end{equation}
which in turn implies $\gamma_0,\gamma_2 \ll 1$, 
since these coefficients belong to different orders in the $f^2/2m^2$ expansion.
The result of Eq.~\eqref{eq:mu2} makes the tuning in the model explicit:
the natural value of $\mu^2$ is not much below the scale $f^2$.
However, once the tuning is implemented,
it is possible to predict the value and the sign of the quartic coupling,
$\lambda(m)$. In fact, since the coefficient 
of $\text{sin}^2(h/f)$ is numerically small,
the leading order for the quartic term, $\frac{1}{4} \lambda(m) h^4$, is given by
the $\text{sin}^4(h/f)$ term:
\begin{equation}\label{eq:lambda0}
 \lambda(m) = \frac{N_c}{8 \pi^2} \frac{f^2}{2 m^2} \delta_2.
\end{equation}
The condition $\gamma_0 \ll 1$ implies
\begin{equation}
 x^2(z_{\mathrm{L}}^2 + z_{\mathrm{R}}^2) \simeq 2 x z_{\mathrm{L}} z_{\mathrm{R}} a_0 
 \cos\Omega\,,
\end{equation}
which then yields
\begin{equation}
 \delta_2 \simeq - 2 x^4 (z_{\mathrm{L}}^2 + z_{\mathrm{R}}^2),
\end{equation}
and finally
\begin{equation}\label{eq:lambdam}
 \lambda(m) = - \frac{N_c}{4 \pi^2} \frac{f^2}{2 m^2} x^4 (z_{\mathrm{L}}^2 
    + z_{\mathrm{R}}^2) < 0.
\end{equation}
Since $\lambda(m)$ is entirely predicted in the SM below the 
threshold of the FN messengers,
Eq.~\eqref{eq:lambdam} can be used to determine the scale $m$ at which
a successful matching is achieved. 
In order to do so, we recall that the goal of the FN mechanism
is to construct a model where 
there is no hierarchy among the fundamental 
parameters.
Since we know that $y_t(m) \simeq x$, the top Yukawa coupling at the scale
$m$ fixes the overall magnitude of the other couplings.
Thus, Eq.~\eqref{eq:lambdam} can be rewritten as
\begin{equation}\label{eq:lambda3}
 \lambda(m) = - \frac{N_c}{ 2 \pi^2} \frac{f^2}{ 2 m^2} 
 y_t^6(m) (1+\delta)^6,
\end{equation}
where we have parametrized an average Yukawa coupling as $y_t(m)(1+\delta)$.

\begin{figure}
\centering
 \includegraphics[scale=0.55]{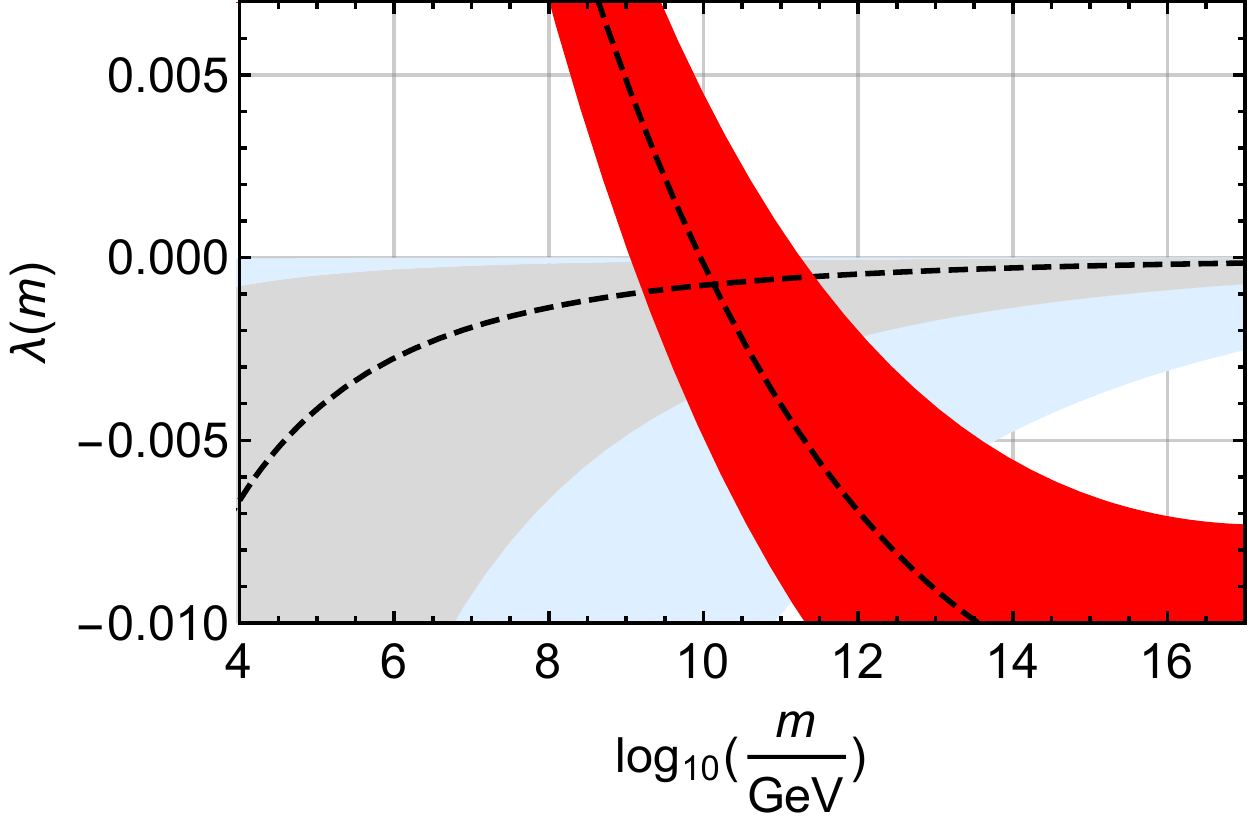}
 \caption{Matching of the Higgs quartic coupling at the scale $m$ 
 in the SM (red band) with the prediction of Eq.~\eqref{eq:lambda3}, considering a Yukawa coupling
 spread of $\delta = \pm 0.6$ (light blue band),
$\delta = \pm 0.3$ (light gray band), and
$\delta = 0$ (dashed black line). The intersection corresponds to the allowed range for $m$. See text for details.}
 \label{plot1}
\end{figure}

In Fig.~\ref{plot1} we show the SM running of $\lambda(m)$ in red
(the band takes into account the uncertainty in the
initial conditions) and the RHS of Eq.~\eqref{eq:lambda3}
for $\delta = \pm 0.6$ (light blue band),
$\delta = \pm 0.3$ (light gray band), and
$\delta = 0$ (dashed black line).
We notice that the matching is possible only for negative values 
of $\lambda(m)$, which selects $10^9 \, \text{GeV}\lesssim m \lesssim 10^{14} \, \text{GeV}$.
By recalling $f^2/ 2 m^2 \simeq 0.23$,
we conclude that
\begin{equation}
 7 \times 10^8 \, \text{GeV} \lesssim f \lesssim 7 \times 10^{13} \, \text{GeV}.
\end{equation}
Since the flavon expectation value, $f$, is related to the axion
decay constant by $f_a = f/N$, 
\begin{equation}
\label{eq:DWN}
 N = \sum_i 2 H_{\Psi_{q_L}^i} - H_{\Psi_{u_R}^i} - H_{\Psi_{d_R}^i} \approx 50,
\end{equation}
the previous bound yields
\begin{equation}\label{eq:bound1fa}
 10^7 \, \text{GeV} \lesssim \, f_a \, \lesssim 
 10^{12} \, \text{GeV}.
\end{equation}

It is useful to confront this region
with constraints following from the flavour-violating couplings of the axiflavon. In fact, limits from searches for the decay $K^+ \to \pi^+ a$ lead to $f_a \gtrsim 7.5 \times 10^{10}$\, GeV at $90\%$\,C.L. \cite{Calibbi:2016hwq}, 
leaving a relatively thin stripe \footnote{Note that the axion couplings
to fermions differ by approximately a factor of two
with respect to the axiflavon case of \cite{Calibbi:2016hwq}, 
which is however cancelled to good approximation by a similar factor entering Eq. \eqref{eq:DWN}.}
of
\begin{equation}
 f_a \approx (10^{11}-10^{12}) \, \text{GeV}.
\end{equation} 
Interestingly, this range will almost entirely be tested by the NA62 experiment, which just started operation \cite{Anelli:2005ju,Fantechi:2014hqa}.

\subsection{Including right-handed neutrinos}

We now discuss the impact of including
right-handed (RH) neutrinos.
Let us consider one family first.
The left-handed doublet $l_{\mathrm{L}}$
and the RH neutrino $N_{\mathrm{R}}$ come 
as $\SO(5)$ spurions (see Eq.~\eqref{eq:spurion}): 
\begin{equation}
\Psi_{\mathrm{L}} = \Delta^T_{\mathrm{L}} l_{\mathrm{L}}, 
    \quad \Psi_N = \Delta_u^T N_{\mathrm{R}}.
\end{equation}
One possibility is to assign flavour charge to $\Psi_N$ such that
the following term is allowed:
\begin{equation}
    \label{eq:majorana}
    \begin{split}
	- \mathcal{L}_N = &\frac{1}{\sqrt{2}} y_N \bar{\Psi}_N \Sigma^\prime {\cal C} \bar{\Psi}_N^T + 
	 \text{h.c.}\\
	 = & -\frac{1}{2} y_N f\, \cos(h/f) \bar{N}_{\mathrm{R}} {\cal C}
	 \bar{N}_{\mathrm{R}}^T e^{\imath a/f} + \text{h.c.},
    \end{split}
\end{equation}
which yields a Majorana mass 
\begin{equation}
    \label{eq:}
    m_{N_{\mathrm{R}}}^2(h) = y_N^2 f^2\,\cos^2(h/f). 
\end{equation}
The Dirac mass term, $m_{\mathrm{D}}$, 
is obtained by integrating out the FN chain:
\begin{equation}
 m_{\mathrm{D}} \sim m_t \epsilon^{\frac{|\delta_\nu|-1}{2}},
\end{equation}
where $\delta_\nu = H_{l_L} - H_{N_{\mathrm{R}}} $.
The light neutrino mass, $m_\nu$, is then given by
\begin{equation}
 m_\nu \sim m_t \epsilon^{|\delta_\nu|-1} \frac{m_t}{m_{N_{\mathrm{R}}}},
\end{equation}
which shows a double suppresion,
originating from the type-I seesaw~\cite{Minkowski:1977sc,GellMannRamondSlansky,Yanagida,Mohapatra:1979ia} and from the FN mechanism.
The impact of Eq.~\eqref{eq:majorana} to the Higgs potential is
\begin{equation}\begin{split}
 \Delta V_\text{AFH}^{(1)} = & -\frac{2}{64 \pi^2} m^4_{N_R}(h)\left( \log
 \frac{m^2_{N_R}(h)}{m^2} - \frac{3}{2}\right) \\
 = & -\frac{1}{32 \pi^2} y_N^4 f^4 \cos^4(h/f) \left[
 \log\left( y_N^2 \frac{f^2}{m^2} \right)\right.\\
  &\left.\hspace{1.5cm}+ \log \left(1 - \sin(h/f)^2\right) - \frac{3}{2} \right] \\
  \simeq &- \frac{1}{16 \pi^2} y_N^4 \left(1 + 
 \log\frac{1}{\tilde{\epsilon}} \right) f^4 \text{sin}^2(h/f) \\
 &+ \frac{1}{32 \pi^2} y_N^4  \log\frac{1}{\tilde{\epsilon}}
 f^4 \text{sin}^4(h/f)\,,
 \end{split}
\end{equation}
where we have defined $(y_N f/m)^2 = \tilde{\epsilon}$.
At the leading order in $f^2/ 2 m^2$, 
the matching conditions now read
\begin{equation}
 \mu^2(m) = \frac{f^2}{16 \pi^2} \left[ 
 2 y_N^4 \left(1 + \log\frac{1}{\tilde{\epsilon}} \right)
 - N_c \gamma_0 
 \right],
\end{equation}
and
\begin{equation}\label{eq:lambdan}
 \lambda(m) = \frac{1}{8 \pi^2} \log
 \frac{1}{\tilde{\epsilon}} \, y_N^4.
\end{equation}
Assuming three almost degenerate 
RH neutrinos, with a typical coupling $y_N$
parametrized as $ y_N = (1 + \delta) y_t$, Eq.~\eqref{eq:lambdan} 
becomes
\begin{equation}\label{eq:lambdan2}
 \lambda(m) =  \frac{3}{8 \pi^2} 
 \log\left(\frac{1}{2 y_t^2(m) (1+\delta)^2 \epsilon}\right)
 (1 + \delta)^4 y_t^4(m).
\end{equation}
\begin{figure}
\centering
 \includegraphics[scale=0.55]{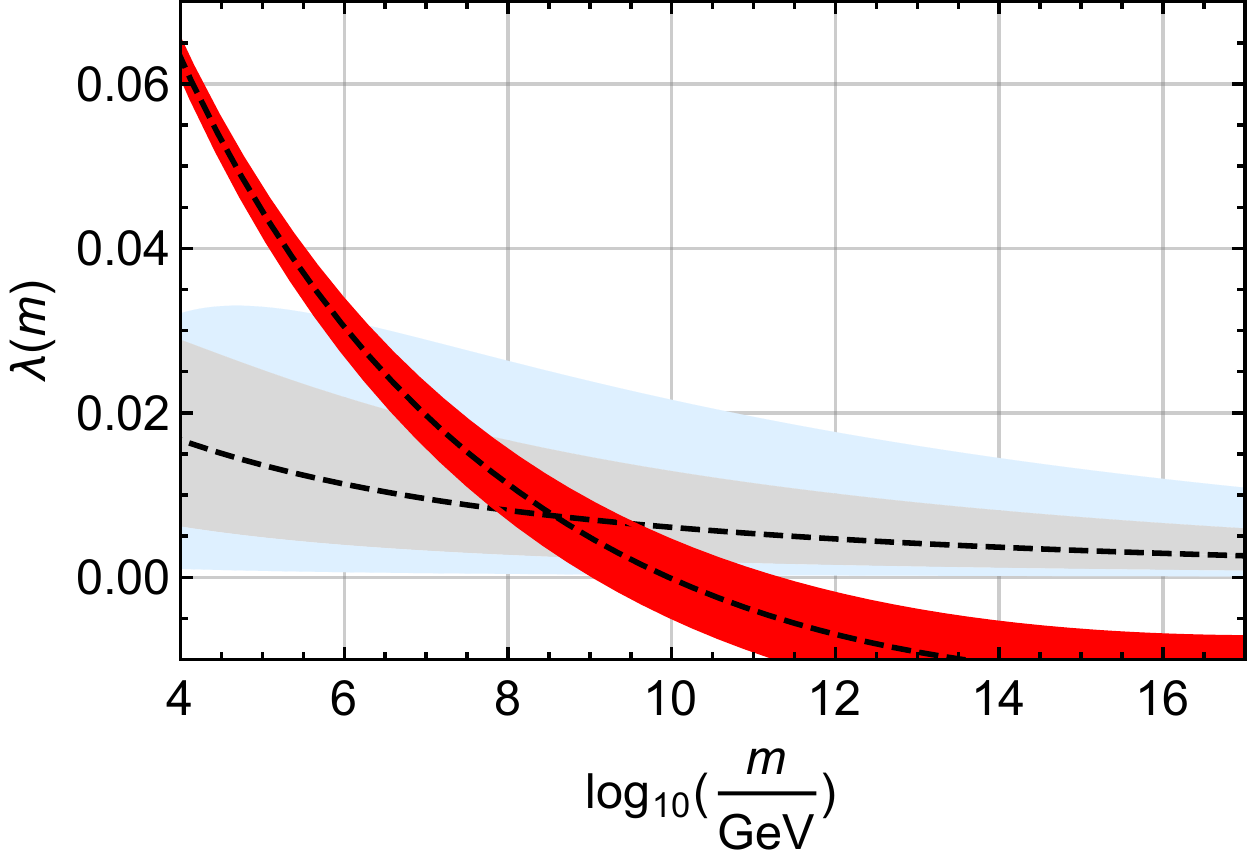}
 \caption{Matching of the Higgs quartic coupling at the scale $m$ in the SM (red band) with the prediction of Eq.\eqref{eq:lambdan2}, considering a yukawa spread of $\delta = \pm 0.6$ (light blue band),
$\delta = \pm 0.3$ (light gray band), and
$\delta = 0$ (dashed black line). The intersection corresponds to the allowed range for $m$. See text for details.}
 \label{plot2}
\end{figure}
In Fig.~\ref{plot2} we show the SM running of $\lambda(m)$ in red
and the RHS of Eq.~\eqref{eq:lambdan2}
for $\delta = \pm 0.6$ (light blue band),
$\delta = \pm 0.3$ (light gray band) and
$\delta = 0$ (dashed black line).
The matching is now possible
for smaller values of $m$ with respect
to the case without RH neutrinos, because the RHS 
of Eq.~\eqref{eq:lambdan2} is positive.
The allowed region for $f$ is:
\begin{equation}\label{eq:RHf}
 3 \times 10^5 \, \text{GeV} \lesssim f \lesssim 10^{11} \text{GeV}.
\end{equation}
Furthermore,
since $m_{N_{\mathrm{R}}} \simeq y_N \, f$ with
$y_N \approx 1$, Eq.~\eqref{eq:RHf} also sets the range of
the RH neutrino masses.
Eventually, we find:
\begin{equation}
 6 \, \text{TeV} \lesssim f_a \lesssim 2 \times 10^{6} \, \text{TeV}\,.
\end{equation}

Such values of $f_a$ are excluded
for the usual QCD axion, however, by disentangling the axion mass and 
decay constant, low-$f_a$ models can 
become viable.
Recent concrete examples have been 
presented in Refs~\cite{Gherghetta:2016fhp,Gaillard:2018xgk}. 
Supernova cooling and flavour
constraints can then be avoided by pushing
the axion mass to the GeV or TeV scale.
As a consequence, the axion cannot be a dark matter candidate, 
since it is no longer 
stable on cosmological scales, but still
solves the strong CP problem. In addition, the RH neutrinos can 
provide a link to 
matter-antimatter asymmetry via leptogenesis~\cite{Alanne:2017sip}.

\subsection{Bounds from inflation}
The inflationary Hubble scale, $H_I$, is constrained by the
CMB measurement of the tensor-to-scalar ratio as
\cite{Planck:2013jfk} 
\begin{equation}\label{eq:HIbound}
 \frac{H_I}{2 \pi} \leq 1.4\times 10^{13}\, \text{GeV}.
\end{equation}
Since the reheating temperature, $T_\text{RH}$, 
is bounded by $H_I$, one obtains  
$ T_\text{RH} \lesssim 10^{16} \, \text{GeV}$
~\cite{Fairbairn:2014zta}.
The expression for $T_\text{RH}$ is model
dependent and is given by~\cite{Chung:1998rq} 
\begin{equation}\label{eq:TRH}
 T_\text{RH} = \left( \frac{ 45 }{4 \pi^3 g_*} \right)^{1/4}
 \left( \Gamma_\phi M_\text{Pl} \right)^{1/2},
\end{equation}
where $\Gamma_\phi$ is the width 
of the inflaton field and $g_*$ is the number
of relativistic degrees of freedom.

Our aim is to extract a cosmological
constraint on the Peccei--Quinn scale when
the flavon field itself plays 
the role of the inflaton, $\phi = \sigma$
\cite{Antusch:2008gw,Fairbairn:2014zta,Ema:2016ops}. 
Note that in order to have a successful 
inflation, one needs, for instance, to introduce
a non-minimal coupling to gravity. 

To estimate the constraint, we compute $\Gamma_\sigma$ in
our model. The flavon field couples
at the tree-level to the other scalars
in the $\Sigma$-multiplet and
to RH neutrinos if the interaction
in Eq.~\eqref{eq:majorana} is included.
The potential in Eq.~\eqref{eq:potential}
is stable for $\lambda_1 > \lambda_2 > 0$.
We focus here in the limit 
where $\lambda_1 \gg \lambda_2$,
such that all the scalar decay channels
for $\sigma$ are open, and the width is thus maximised. 
The width into scalars is then given by
\begin{equation}
 \Gamma_\sigma^{(s)} = N_s \frac{1}{8 \pi^2} 
 \frac{\lambda_1}{2} m_\sigma = N_s \frac{1}{8 \pi^2}
 (\lambda_1)^{3/2} f\,,
\end{equation}
where $N_s = 9$ is the number of scalar decay channels, 
corresponding to two scalar doublets and the axion. 
Including RH neutrinos, which contribute
\begin{equation}
    \begin{split}
	 \Gamma_\sigma^{(N_{\mathrm{R}})} = \frac{3}{32 \pi^2}
	 y_N^2 m_\sigma
	 = \frac{3}{16 \pi^2} (\lambda_1)^{1/2} y_N^2 f\,,
    \end{split}
\end{equation}
we arrive at an upper bound on $f$ of
\begin{equation}
    \begin{split}
	f_a \cdot N = f \leq &\ \frac{ 8 \pi^2}{\left( \lambda_1)^{1/2}(N_s \lambda_1 
	 + \frac{1}{2} \sum_i g_i^2 \right)} 
	 \left( \frac{4 \pi^3 g_*}{45} \right)^{1/2}\\
	 &\times\left( \frac{ 10^{16} \, \text{GeV}}{M_\text{Pl}} \right)
	 \, 10^{16} \, \text{GeV}.
    \end{split}
\end{equation}
This leads to a relevant bound only in the high-scale model without RH neutrinos,
since the one with RH neutrinos easily avoids this constraint due to
the relatively low scale $f$.
Taking $g_* = \mathcal{O}(100)$ and $y_N = 0$, 
we find
\begin{equation}
 f_a \lesssim 3 \, (\lambda_1)^{-3/2} \, 10^{13} \, \text{GeV}\,.
\end{equation}
In the extreme case of $\lambda_1 = 4 \pi$
(for which $\Gamma_\sigma$ almost equals $m_\sigma$),
$f_a \lesssim 10^{12} \, \text{GeV}$, which is
the same upper bound as in Eq.~\eqref{eq:bound1fa}.

\section{Conclusions}
\label{sec:con}

We presented a model framework where the flavour puzzle, the strong CP problem, and the
origin of the observed DM abundance are solved by a single scalar multiplet which also contains the Higgs boson. The latter emerges as
a pNGB of an enhanced global symmetry, thereby connecting the EWSB with the origin of the SM-fermion mass hierarchies. 
To achieve this, we provided a renormalizable UV realization of the FN mechanism.

We showed that successfully reproducing the SM-like Higgs potential at low energies 
fixes the axion decay constant to $f_a \approx (10^{11}-10^{12}) \, \text{GeV}$ for the minimal setup, 
where the lower bound is driven from limits on flavour-changing axion couplings. 
This lies just in a region where axions are a very attractive
DM candidate, and which is a prime target of
future searches, in particular at NA62 and ADMX~\cite{Asztalos:2009yp,Du:2018uak}.

We also demonstrated that by including RH neutrinos, the symmetry-breaking scale could
be lowered bringing the axion decay constant down to TeV range and commented on the possibility of the 
flavon component to be identified with the inflaton.

\section*{Acknowledgments}
We are grateful to Giorgio Arcadi, Pablo Quilez, Kai Schmitz, Stefan Vogl and Kei Yagyu for useful discussions.

\appendix

\section{The validity of the FN mass degeneracy assumption}

\label{app:m}

Let us discuss the implication of relaxing the degeneracy
among the FN messengers. We paramatrize
the departure from the degenerate case by introducing
$\delta_j$ parameters:
\begin{equation}
    \begin{split}
	 m^2_{\xi_j} &= m^2(1 + \delta_j) + f_j(h) =
	 m^2 + (m^2 \delta_j + f_j(h)) \\
	 &\equiv m^2 + \tilde{f}_j(h).
    \end{split}
\end{equation}
The previous analysis now applies with $f_j(h)$ replaced by
$\tilde{f}_j(h)$.
The condition under which the non-degeneracy
effects can be neglected then reads:
\begin{equation}
 m^2 \delta_j \ll f \times m \Rightarrow
 \delta_j \ll f/m =\sqrt{2\epsilon}\sim 0.68
\end{equation}
We conclude that the non-degeneracy
can be neglected as long as
$\delta_j \ll 0.7$.

\section{Subleading contributions to the Higgs potential}

\label{app:sub}
Similarly to the top eigenstate, the contribution of
the gauge bosons will appear on both sides of Eq.~\eqref{eq:match1}
and thus do not alter the one-loop
matching condition;
the same reasoning applies to the other SM eigenstates.
However, an additional contribution
is expected when the FN messengers mixing with the light
fermions are included.

For concreteness, let us consider the up-type quarks.
The Yukawa interactions before diagonalization read 
\begin{equation}\label{eq:Lupquarks}
\mathcal{L}_u = - y_{ij} \epsilon^{n_{ij}} \bar{q}_L^{\,i} \tilde{H} u_R^{\,j} +{\rm h.c.}\,,
\end{equation}
where $y_{ij} = \mathcal{O}(1)$, where the hierarchies
\begin{equation}\label{eq:nij}
 n_{ij} = \left( \begin{array}{ccc}
           8 & 4 & 3 \\
           7 & 3 & 2 \\
           5 & 1 & 0
          \end{array}\right)
\end{equation}
lead to a viable spectrum~\cite{Ema:2016ops}.
Thus, by recalling the relation between $\UH$ charge differences
and the suppresion factor Eq.~\eqref{eq:suppr}, we shall require
\begin{equation}
 \frac{|\delta_{ij}| - 1}{2} = n_{ij}.
\end{equation}
Looking at the second-third family
mixing, we see that it can be reproduced by assigning the $H$ charges
\begin{equation}\label{eq:charmcharge}
 H_{{t_{\mathrm{L}}}} = 1,\ H_{{t_{\mathrm{R}}}} = 2,\ H_{{c_{\mathrm{L}}}} = -3,\ H_{{c_{\mathrm{R}}}} = 4.
\end{equation}
The charm sector can then be written in a similar
way as in Eq.~\eqref{eq:xicsi}, according to the
charge assignments, Eq.~\eqref{eq:charmcharge}.
The contribution to the Higgs potential
is found to be 
$\mathcal{O}(f^2/ 2 m^2)$ for the $\text{sin}^2(h/f)$
and $\mathcal{O}(f^4/ 4 m^4)$ for the $\text{sin}^4(h/f)$ term,
which is subleading compared to the top sector.
The same reasoning applies to all light SM fermions.

\bibliography{biblio.bib}

\begin{thebibliography}{33}%
\makeatletter
\providecommand \@ifxundefined [1]{%
 \@ifx{#1\undefined}
}%
\providecommand \@ifnum [1]{%
 \ifnum #1\expandafter \@firstoftwo
 \else \expandafter \@secondoftwo
 \fi
}%
\providecommand \@ifx [1]{%
 \ifx #1\expandafter \@firstoftwo
 \else \expandafter \@secondoftwo
 \fi
}%
\providecommand \natexlab [1]{#1}%
\providecommand \enquote  [1]{``#1''}%
\providecommand \bibnamefont  [1]{#1}%
\providecommand \bibfnamefont [1]{#1}%
\providecommand \citenamefont [1]{#1}%
\providecommand \href@noop [0]{\@secondoftwo}%
\providecommand \href [0]{\begingroup \@sanitize@url \@href}%
\providecommand \@href[1]{\@@startlink{#1}\@@href}%
\providecommand \@@href[1]{\endgroup#1\@@endlink}%
\providecommand \@sanitize@url [0]{\catcode `\\12\catcode `\$12\catcode
  `\&12\catcode `\#12\catcode `\^12\catcode `\_12\catcode `\%12\relax}%
\providecommand \@@startlink[1]{}%
\providecommand \@@endlink[0]{}%
\providecommand \url  [0]{\begingroup\@sanitize@url \@url }%
\providecommand \@url [1]{\endgroup\@href {#1}{\urlprefix }}%
\providecommand \urlprefix  [0]{URL }%
\providecommand \Eprint [0]{\href }%
\providecommand \doibase [0]{http://dx.doi.org/}%
\providecommand \selectlanguage [0]{\@gobble}%
\providecommand \bibinfo  [0]{\@secondoftwo}%
\providecommand \bibfield  [0]{\@secondoftwo}%
\providecommand \translation [1]{[#1]}%
\providecommand \BibitemOpen [0]{}%
\providecommand \bibitemStop [0]{}%
\providecommand \bibitemNoStop [0]{.\EOS\space}%
\providecommand \EOS [0]{\spacefactor3000\relax}%
\providecommand \BibitemShut  [1]{\csname bibitem#1\endcsname}%
\let\auto@bib@innerbib\@empty
\bibitem [{\citenamefont {Froggatt}\ and\ \citenamefont
  {Nielsen}(1979)}]{Froggatt:1978nt}%
  \BibitemOpen
  \bibfield  {author} {\bibinfo {author} {\bibfnamefont {C.~D.}\ \bibnamefont
  {Froggatt}}\ and\ \bibinfo {author} {\bibfnamefont {H.~B.}\ \bibnamefont
  {Nielsen}},\ }\href {\doibase 10.1016/0550-3213(79)90316-X} {\bibfield
  {journal} {\bibinfo  {journal} {Nucl. Phys.}\ }\textbf {\bibinfo {volume}
  {B147}},\ \bibinfo {pages} {277} (\bibinfo {year} {1979})}\BibitemShut
  {NoStop}%
\bibitem [{\citenamefont {Calibbi}\ \emph {et~al.}(2017)\citenamefont
  {Calibbi}, \citenamefont {Goertz}, \citenamefont {Redigolo}, \citenamefont
  {Ziegler},\ and\ \citenamefont {Zupan}}]{Calibbi:2016hwq}%
  \BibitemOpen
  \bibfield  {author} {\bibinfo {author} {\bibfnamefont {L.}~\bibnamefont
  {Calibbi}}, \bibinfo {author} {\bibfnamefont {F.}~\bibnamefont {Goertz}},
  \bibinfo {author} {\bibfnamefont {D.}~\bibnamefont {Redigolo}}, \bibinfo
  {author} {\bibfnamefont {R.}~\bibnamefont {Ziegler}}, \ and\ \bibinfo
  {author} {\bibfnamefont {J.}~\bibnamefont {Zupan}},\ }\href {\doibase
  10.1103/PhysRevD.95.095009} {\bibfield  {journal} {\bibinfo  {journal} {Phys.
  Rev.}\ }\textbf {\bibinfo {volume} {D95}},\ \bibinfo {pages} {095009}
  (\bibinfo {year} {2017})},\ \Eprint {http://arxiv.org/abs/1612.08040}
  {arXiv:1612.08040 [hep-ph]} \BibitemShut {NoStop}%
\bibitem [{\citenamefont {Ema}\ \emph {et~al.}(2017)\citenamefont {Ema},
  \citenamefont {Hamaguchi}, \citenamefont {Moroi},\ and\ \citenamefont
  {Nakayama}}]{Ema:2016ops}%
  \BibitemOpen
  \bibfield  {author} {\bibinfo {author} {\bibfnamefont {Y.}~\bibnamefont
  {Ema}}, \bibinfo {author} {\bibfnamefont {K.}~\bibnamefont {Hamaguchi}},
  \bibinfo {author} {\bibfnamefont {T.}~\bibnamefont {Moroi}}, \ and\ \bibinfo
  {author} {\bibfnamefont {K.}~\bibnamefont {Nakayama}},\ }\href {\doibase
  10.1007/JHEP01(2017)096} {\bibfield  {journal} {\bibinfo  {journal} {JHEP}\
  }\textbf {\bibinfo {volume} {01}},\ \bibinfo {pages} {096} (\bibinfo {year}
  {2017})},\ \Eprint {http://arxiv.org/abs/1612.05492} {arXiv:1612.05492
  [hep-ph]} \BibitemShut {NoStop}%
\bibitem [{\citenamefont {Peccei}\ and\ \citenamefont
  {Quinn}(1977)}]{Peccei:1977hh}%
  \BibitemOpen
  \bibfield  {author} {\bibinfo {author} {\bibfnamefont {R.~D.}\ \bibnamefont
  {Peccei}}\ and\ \bibinfo {author} {\bibfnamefont {H.~R.}\ \bibnamefont
  {Quinn}},\ }\href {\doibase 10.1103/PhysRevLett.38.1440} {\bibfield
  {journal} {\bibinfo  {journal} {Phys. Rev. Lett.}\ }\textbf {\bibinfo
  {volume} {38}},\ \bibinfo {pages} {1440} (\bibinfo {year} {1977})},\ \bibinfo
  {note} {[,328(1977)]}\BibitemShut {NoStop}%
\bibitem [{\citenamefont {Wilczek}(1978)}]{Wilczek:1977pj}%
  \BibitemOpen
  \bibfield  {author} {\bibinfo {author} {\bibfnamefont {F.}~\bibnamefont
  {Wilczek}},\ }\href {\doibase 10.1103/PhysRevLett.40.279} {\bibfield
  {journal} {\bibinfo  {journal} {Phys. Rev. Lett.}\ }\textbf {\bibinfo
  {volume} {40}},\ \bibinfo {pages} {279} (\bibinfo {year} {1978})}\BibitemShut
  {NoStop}%
\bibitem [{\citenamefont {Weinberg}(1978)}]{Weinberg:1977ma}%
  \BibitemOpen
  \bibfield  {author} {\bibinfo {author} {\bibfnamefont {S.}~\bibnamefont
  {Weinberg}},\ }\href {\doibase 10.1103/PhysRevLett.40.223} {\bibfield
  {journal} {\bibinfo  {journal} {Phys. Rev. Lett.}\ }\textbf {\bibinfo
  {volume} {40}},\ \bibinfo {pages} {223} (\bibinfo {year} {1978})}\BibitemShut
  {NoStop}%
\bibitem [{\citenamefont {Kim}(1979)}]{Kim:1979if}%
  \BibitemOpen
  \bibfield  {author} {\bibinfo {author} {\bibfnamefont {J.~E.}\ \bibnamefont
  {Kim}},\ }\href {\doibase 10.1103/PhysRevLett.43.103} {\bibfield  {journal}
  {\bibinfo  {journal} {Phys. Rev. Lett.}\ }\textbf {\bibinfo {volume} {43}},\
  \bibinfo {pages} {103} (\bibinfo {year} {1979})}\BibitemShut {NoStop}%
\bibitem [{\citenamefont {Shifman}\ \emph {et~al.}(1980)\citenamefont
  {Shifman}, \citenamefont {Vainshtein},\ and\ \citenamefont
  {Zakharov}}]{Shifman:1979if}%
  \BibitemOpen
  \bibfield  {author} {\bibinfo {author} {\bibfnamefont {M.~A.}\ \bibnamefont
  {Shifman}}, \bibinfo {author} {\bibfnamefont {A.~I.}\ \bibnamefont
  {Vainshtein}}, \ and\ \bibinfo {author} {\bibfnamefont {V.~I.}\ \bibnamefont
  {Zakharov}},\ }\href {\doibase 10.1016/0550-3213(80)90209-6} {\bibfield
  {journal} {\bibinfo  {journal} {Nucl. Phys.}\ }\textbf {\bibinfo {volume}
  {B166}},\ \bibinfo {pages} {493} (\bibinfo {year} {1980})}\BibitemShut
  {NoStop}%
\bibitem [{\citenamefont {Dine}\ \emph {et~al.}(1981)\citenamefont {Dine},
  \citenamefont {Fischler},\ and\ \citenamefont {Srednicki}}]{Dine:1981rt}%
  \BibitemOpen
  \bibfield  {author} {\bibinfo {author} {\bibfnamefont {M.}~\bibnamefont
  {Dine}}, \bibinfo {author} {\bibfnamefont {W.}~\bibnamefont {Fischler}}, \
  and\ \bibinfo {author} {\bibfnamefont {M.}~\bibnamefont {Srednicki}},\ }\href
  {\doibase 10.1016/0370-2693(81)90590-6} {\bibfield  {journal} {\bibinfo
  {journal} {Phys. Lett.}\ }\textbf {\bibinfo {volume} {104B}},\ \bibinfo
  {pages} {199} (\bibinfo {year} {1981})}\BibitemShut {NoStop}%
\bibitem [{\citenamefont {Zhitnitsky}(1980)}]{Zhitnitsky:1980tq}%
  \BibitemOpen
  \bibfield  {author} {\bibinfo {author} {\bibfnamefont {A.~R.}\ \bibnamefont
  {Zhitnitsky}},\ }\href@noop {} {\bibfield  {journal} {\bibinfo  {journal}
  {Sov. J. Nucl. Phys.}\ }\textbf {\bibinfo {volume} {31}},\ \bibinfo {pages}
  {260} (\bibinfo {year} {1980})},\ \bibinfo {note} {[Yad.
  Fiz.31,497(1980)]}\BibitemShut {NoStop}%
\bibitem [{\citenamefont {Preskill}\ \emph {et~al.}(1983)\citenamefont
  {Preskill}, \citenamefont {Wise},\ and\ \citenamefont
  {Wilczek}}]{Preskill:1982cy}%
  \BibitemOpen
  \bibfield  {author} {\bibinfo {author} {\bibfnamefont {J.}~\bibnamefont
  {Preskill}}, \bibinfo {author} {\bibfnamefont {M.~B.}\ \bibnamefont {Wise}},
  \ and\ \bibinfo {author} {\bibfnamefont {F.}~\bibnamefont {Wilczek}},\ }\href
  {\doibase 10.1016/0370-2693(83)90637-8} {\bibfield  {journal} {\bibinfo
  {journal} {Phys. Lett.}\ }\textbf {\bibinfo {volume} {B120}},\ \bibinfo
  {pages} {127} (\bibinfo {year} {1983})},\ \BibitemShut {NoStop}%
\bibitem [{\citenamefont {Abbott}\ and\ \citenamefont
  {Sikivie}(1983)}]{Abbott:1982af}%
  \BibitemOpen
  \bibfield  {author} {\bibinfo {author} {\bibfnamefont {L.~F.}\ \bibnamefont
  {Abbott}}\ and\ \bibinfo {author} {\bibfnamefont {P.}~\bibnamefont
  {Sikivie}},\ }\href {\doibase 10.1016/0370-2693(83)90638-X} {\bibfield
  {journal} {\bibinfo  {journal} {Phys. Lett.}\ }\textbf {\bibinfo {volume}
  {B120}},\ \bibinfo {pages} {133} (\bibinfo {year} {1983})},\ \BibitemShut {NoStop}%
\bibitem [{\citenamefont {Dine}\ and\ \citenamefont
  {Fischler}(1983)}]{Dine:1982ah}%
  \BibitemOpen
  \bibfield  {author} {\bibinfo {author} {\bibfnamefont {M.}~\bibnamefont
  {Dine}}\ and\ \bibinfo {author} {\bibfnamefont {W.}~\bibnamefont
  {Fischler}},\ }\href {\doibase 10.1016/0370-2693(83)90639-1} {\bibfield
  {journal} {\bibinfo  {journal} {Phys. Lett.}\ }\textbf {\bibinfo {volume}
  {B120}},\ \bibinfo {pages} {137} (\bibinfo {year} {1983})},\ \BibitemShut {NoStop}%
\bibitem [{\citenamefont {Redi}\ and\ \citenamefont
  {Strumia}(2012)}]{Redi:2012ad}%
  \BibitemOpen
  \bibfield  {author} {\bibinfo {author} {\bibfnamefont {M.}~\bibnamefont
  {Redi}}\ and\ \bibinfo {author} {\bibfnamefont {A.}~\bibnamefont {Strumia}},\
  }\href {\doibase 10.1007/JHEP11(2012)103} {\bibfield  {journal} {\bibinfo
  {journal} {JHEP}\ }\textbf {\bibinfo {volume} {11}},\ \bibinfo {pages} {103}
  (\bibinfo {year} {2012})},\ \Eprint {http://arxiv.org/abs/1208.6013}
  {arXiv:1208.6013 [hep-ph]} \BibitemShut {NoStop}%
\bibitem [{\citenamefont {Alanne}\ \emph {et~al.}(2015)\citenamefont {Alanne},
  \citenamefont {Gertov}, \citenamefont {Sannino},\ and\ \citenamefont
  {Tuominen}}]{Alanne:2014kea}%
  \BibitemOpen
  \bibfield  {author} {\bibinfo {author} {\bibfnamefont {T.}~\bibnamefont
  {Alanne}}, \bibinfo {author} {\bibfnamefont {H.}~\bibnamefont {Gertov}},
  \bibinfo {author} {\bibfnamefont {F.}~\bibnamefont {Sannino}}, \ and\
  \bibinfo {author} {\bibfnamefont {K.}~\bibnamefont {Tuominen}},\ }\href
  {\doibase 10.1103/PhysRevD.91.095021} {\bibfield  {journal} {\bibinfo
  {journal} {Phys. Rev.}\ }\textbf {\bibinfo {volume} {D91}},\ \bibinfo {pages}
  {095021} (\bibinfo {year} {2015})},\ \Eprint {http://arxiv.org/abs/1411.6132}
  {arXiv:1411.6132 [hep-ph]} \BibitemShut {NoStop}%
\bibitem [{\citenamefont {Alanne}\ \emph {et~al.}(2016)\citenamefont {Alanne},
  \citenamefont {Meroni}, \citenamefont {Sannino},\ and\ \citenamefont
  {Tuominen}}]{Alanne:2015fqh}%
  \BibitemOpen
  \bibfield  {author} {\bibinfo {author} {\bibfnamefont {T.}~\bibnamefont
  {Alanne}}, \bibinfo {author} {\bibfnamefont {A.}~\bibnamefont {Meroni}},
  \bibinfo {author} {\bibfnamefont {F.}~\bibnamefont {Sannino}}, \ and\
  \bibinfo {author} {\bibfnamefont {K.}~\bibnamefont {Tuominen}},\ }\href
  {\doibase 10.1103/PhysRevD.93.091701} {\bibfield  {journal} {\bibinfo
  {journal} {Phys. Rev.}\ }\textbf {\bibinfo {volume} {D93}},\ \bibinfo {pages}
  {091701} (\bibinfo {year} {2016})},\ \Eprint
  {http://arxiv.org/abs/1511.01910} {arXiv:1511.01910 [hep-ph]} \BibitemShut
  {NoStop}%
\bibitem [{\citenamefont {Kaplan}(1991)}]{Kaplan:1991dc}%
  \BibitemOpen
  \bibfield  {author} {\bibinfo {author} {\bibfnamefont {D.~B.}\ \bibnamefont
  {Kaplan}},\ }\href {\doibase 10.1016/S0550-3213(05)80021-5} {\bibfield
  {journal} {\bibinfo  {journal} {Nucl. Phys.}\ }\textbf {\bibinfo {volume}
  {B365}},\ \bibinfo {pages} {259} (\bibinfo {year} {1991})}\BibitemShut
  {NoStop}%
\bibitem [{\citenamefont {Agashe}\ \emph {et~al.}(2005)\citenamefont {Agashe},
  \citenamefont {Contino},\ and\ \citenamefont {Pomarol}}]{Agashe:2004rs}%
  \BibitemOpen
  \bibfield  {author} {\bibinfo {author} {\bibfnamefont {K.}~\bibnamefont
  {Agashe}}, \bibinfo {author} {\bibfnamefont {R.}~\bibnamefont {Contino}}, \
  and\ \bibinfo {author} {\bibfnamefont {A.}~\bibnamefont {Pomarol}},\ }\href
  {\doibase 10.1016/j.nuclphysb.2005.04.035} {\bibfield  {journal} {\bibinfo
  {journal} {Nucl. Phys.}\ }\textbf {\bibinfo {volume} {B719}},\ \bibinfo
  {pages} {165} (\bibinfo {year} {2005})},\ \Eprint
  {http://arxiv.org/abs/hep-ph/0412089} {arXiv:hep-ph/0412089 [hep-ph]}
  \BibitemShut {NoStop}%
\bibitem [{\citenamefont {Anelli}\ \emph {et~al.}(2005)\citenamefont {Anelli}
  \emph {et~al.}}]{Anelli:2005ju}%
  \BibitemOpen
  \bibfield  {author} {\bibinfo {author} {\bibfnamefont {G.}~\bibnamefont
  {Anelli}} \emph {et~al.},\ }\href@noop {} {\bibfield  {journal} {\bibinfo
  {journal} {CERN-SPSC-2005-013, CERN-SPSC-P-326}\ } (\bibinfo {year}
  {2005})}\BibitemShut {NoStop}%
\bibitem [{\citenamefont {Fantechi}(2014)}]{Fantechi:2014hqa}%
  \BibitemOpen
  \bibfield  {author} {\bibinfo {author} {\bibfnamefont {R.}~\bibnamefont
  {Fantechi}} (\bibinfo {collaboration} {NA62}),\ }\href
  {https://inspirehep.net/record/1309159/files/arXiv:1407.8213.pdf} {\bibfield
  {journal} {\bibinfo  {journal} {in 12th Conference on Flavor Physics and CP
  Violation (FPCP 2014) Marseille, France, May 26-30, 2014}\ } (\bibinfo {year}
  {2014})},\ \Eprint {http://arxiv.org/abs/1407.8213} {arXiv:1407.8213
  [physics.ins-det]} \BibitemShut {NoStop}%
\bibitem [{\citenamefont {Minkowski}(1977)}]{Minkowski:1977sc}%
  \BibitemOpen
  \bibfield  {author} {\bibinfo {author} {\bibfnamefont {P.}~\bibnamefont
  {Minkowski}},\ }\href {\doibase 10.1016/0370-2693(77)90435-X} {\bibfield
  {journal} {\bibinfo  {journal} {Phys. Lett.}\ }\textbf {\bibinfo {volume}
  {67B}},\ \bibinfo {pages} {421} (\bibinfo {year} {1977})}\BibitemShut
  {NoStop}%
\bibitem [{\citenamefont {Gell-Mann}\ \emph {et~al.}(1979)\citenamefont
  {Gell-Mann}, \citenamefont {Ramond},\ and\ \citenamefont
  {Slansky}}]{GellMannRamondSlansky}%
  \BibitemOpen
  \bibfield  {author} {\bibinfo {author} {\bibfnamefont {M.}~\bibnamefont
  {Gell-Mann}}, \bibinfo {author} {\bibfnamefont {P.}~\bibnamefont {Ramond}}, \
  and\ \bibinfo {author} {\bibfnamefont {R.}~\bibnamefont {Slansky}},\
  }\href@noop {} {\emph {\bibinfo {title} {{Supergravity}}}},\ edited by\
  \bibinfo {editor} {\bibnamefont {{F. Nieuwenhuizen and D. Friedman}}}\
  (\bibinfo  {publisher} {{North Holland, Amsterdam}},\ \bibinfo {year}
  {{1979}})\ p.\ \bibinfo {pages} {315}\BibitemShut {NoStop}%
\bibitem [{\citenamefont {{T. Yanagida}}(1979)}]{Yanagida}%
  \BibitemOpen
  \bibfield  {author} {\bibinfo {author} {\bibnamefont {{T. Yanagida}}},\
  }\href@noop {} {\enquote {\bibinfo {title} {{Proc. of the Workshop on Unified
  Theories and the Baryon Number of the Universe}},}\ }\bibinfo {howpublished}
  {{KEK, Japan}} (\bibinfo {year} {{1979}})\BibitemShut {NoStop}%
\bibitem [{\citenamefont {Mohapatra}\ and\ \citenamefont
  {Senjanovic}(1980)}]{Mohapatra:1979ia}%
  \BibitemOpen
  \bibfield  {author} {\bibinfo {author} {\bibfnamefont {R.~N.}\ \bibnamefont
  {Mohapatra}}\ and\ \bibinfo {author} {\bibfnamefont {G.}~\bibnamefont
  {Senjanovic}},\ }\href {\doibase 10.1103/PhysRevLett.44.912} {\bibfield
  {journal} {\bibinfo  {journal} {Phys. Rev. Lett.}\ }\textbf {\bibinfo
  {volume} {44}},\ \bibinfo {pages} {912} (\bibinfo {year} {1980})}\BibitemShut
  {NoStop}%
\bibitem [{\citenamefont {Gherghetta}\ \emph {et~al.}(2016)\citenamefont
  {Gherghetta}, \citenamefont {Nagata},\ and\ \citenamefont
  {Shifman}}]{Gherghetta:2016fhp}%
  \BibitemOpen
  \bibfield  {author} {\bibinfo {author} {\bibfnamefont {T.}~\bibnamefont
  {Gherghetta}}, \bibinfo {author} {\bibfnamefont {N.}~\bibnamefont {Nagata}},
  \ and\ \bibinfo {author} {\bibfnamefont {M.}~\bibnamefont {Shifman}},\ }\href
  {\doibase 10.1103/PhysRevD.93.115010} {\bibfield  {journal} {\bibinfo
  {journal} {Phys. Rev.}\ }\textbf {\bibinfo {volume} {D93}},\ \bibinfo {pages}
  {115010} (\bibinfo {year} {2016})},\ \Eprint
  {http://arxiv.org/abs/1604.01127} {arXiv:1604.01127 [hep-ph]} \BibitemShut
  {NoStop}%
\bibitem [{\citenamefont {Gaillard}\ \emph {et~al.}(2018)\citenamefont
  {Gaillard}, \citenamefont {Gavela}, \citenamefont {Houtz}, \citenamefont
  {Quilez},\ and\ \citenamefont {Del~Rey}}]{Gaillard:2018xgk}%
  \BibitemOpen
  \bibfield  {author} {\bibinfo {author} {\bibfnamefont {M.~K.}\ \bibnamefont
  {Gaillard}}, \bibinfo {author} {\bibfnamefont {M.~B.}\ \bibnamefont
  {Gavela}}, \bibinfo {author} {\bibfnamefont {R.}~\bibnamefont {Houtz}},
  \bibinfo {author} {\bibfnamefont {P.}~\bibnamefont {Quilez}}, \ and\ \bibinfo
  {author} {\bibfnamefont {R.}~\bibnamefont {Del~Rey}},\ }\href@noop {} {\
  (\bibinfo {year} {2018})},\ \Eprint {http://arxiv.org/abs/1805.06465}
  {arXiv:1805.06465 [hep-ph]} \BibitemShut {NoStop}%
\bibitem [{\citenamefont {Alanne}\ \emph {et~al.}(2017)\citenamefont {Alanne},
  \citenamefont {Meroni},\ and\ \citenamefont {Tuominen}}]{Alanne:2017sip}%
  \BibitemOpen
  \bibfield  {author} {\bibinfo {author} {\bibfnamefont {T.}~\bibnamefont
  {Alanne}}, \bibinfo {author} {\bibfnamefont {A.}~\bibnamefont {Meroni}}, \
  and\ \bibinfo {author} {\bibfnamefont {K.}~\bibnamefont {Tuominen}},\ }\href
  {\doibase 10.1103/PhysRevD.96.095015} {\bibfield  {journal} {\bibinfo
  {journal} {Phys. Rev.}\ }\textbf {\bibinfo {volume} {D96}},\ \bibinfo {pages}
  {095015} (\bibinfo {year} {2017})},\ \Eprint
  {http://arxiv.org/abs/1706.10128} {arXiv:1706.10128 [hep-ph]} \BibitemShut
  {NoStop}%
\bibitem [{\citenamefont {Ade}\ \emph {et~al.}(2014)\citenamefont {Ade} \emph
  {et~al.}}]{Planck:2013jfk}%
  \BibitemOpen
  \bibfield  {author} {\bibinfo {author} {\bibfnamefont {P.~A.~R.}\
  \bibnamefont {Ade}} \emph {et~al.} (\bibinfo {collaboration} {Planck}),\
  }\href {\doibase 10.1051/0004-6361/201321569} {\bibfield  {journal} {\bibinfo
   {journal} {Astron. Astrophys.}\ }\textbf {\bibinfo {volume} {571}},\
  \bibinfo {pages} {A22} (\bibinfo {year} {2014})},\ \Eprint
  {http://arxiv.org/abs/1303.5082} {arXiv:1303.5082 [astro-ph.CO]} \BibitemShut
  {NoStop}%
\bibitem [{\citenamefont {Fairbairn}\ \emph {et~al.}(2015)\citenamefont
  {Fairbairn}, \citenamefont {Hogan},\ and\ \citenamefont
  {Marsh}}]{Fairbairn:2014zta}%
  \BibitemOpen
  \bibfield  {author} {\bibinfo {author} {\bibfnamefont {M.}~\bibnamefont
  {Fairbairn}}, \bibinfo {author} {\bibfnamefont {R.}~\bibnamefont {Hogan}}, \
  and\ \bibinfo {author} {\bibfnamefont {D.~J.~E.}\ \bibnamefont {Marsh}},\
  }\href {\doibase 10.1103/PhysRevD.91.023509} {\bibfield  {journal} {\bibinfo
  {journal} {Phys. Rev.}\ }\textbf {\bibinfo {volume} {D91}},\ \bibinfo {pages}
  {023509} (\bibinfo {year} {2015})},\ \Eprint {http://arxiv.org/abs/1410.1752}
  {arXiv:1410.1752 [hep-ph]} \BibitemShut {NoStop}%
\bibitem [{\citenamefont {Chung}\ \emph {et~al.}(1999)\citenamefont {Chung},
  \citenamefont {Kolb},\ and\ \citenamefont {Riotto}}]{Chung:1998rq}%
  \BibitemOpen
  \bibfield  {author} {\bibinfo {author} {\bibfnamefont {D.~J.~H.}\
  \bibnamefont {Chung}}, \bibinfo {author} {\bibfnamefont {E.~W.}\ \bibnamefont
  {Kolb}}, \ and\ \bibinfo {author} {\bibfnamefont {A.}~\bibnamefont
  {Riotto}},\ }\href {\doibase 10.1103/PhysRevD.60.063504} {\bibfield
  {journal} {\bibinfo  {journal} {Phys. Rev.}\ }\textbf {\bibinfo {volume}
  {D60}},\ \bibinfo {pages} {063504} (\bibinfo {year} {1999})},\ \Eprint
  {http://arxiv.org/abs/hep-ph/9809453} {arXiv:hep-ph/9809453 [hep-ph]}
  \BibitemShut {NoStop}%
\bibitem [{\citenamefont {Antusch}\ \emph {et~al.}(2008)\citenamefont
  {Antusch}, \citenamefont {King}, \citenamefont {Malinsky}, \citenamefont
  {Velasco-Sevilla},\ and\ \citenamefont {Zavala}}]{Antusch:2008gw}%
  \BibitemOpen
  \bibfield  {author} {\bibinfo {author} {\bibfnamefont {S.}~\bibnamefont
  {Antusch}}, \bibinfo {author} {\bibfnamefont {S.~F.}\ \bibnamefont {King}},
  \bibinfo {author} {\bibfnamefont {M.}~\bibnamefont {Malinsky}}, \bibinfo
  {author} {\bibfnamefont {L.}~\bibnamefont {Velasco-Sevilla}}, \ and\ \bibinfo
  {author} {\bibfnamefont {I.}~\bibnamefont {Zavala}},\ }\href {\doibase
  10.1016/j.physletb.2008.07.051} {\bibfield  {journal} {\bibinfo  {journal}
  {Phys. Lett.}\ }\textbf {\bibinfo {volume} {B666}},\ \bibinfo {pages} {176}
  (\bibinfo {year} {2008})},\ \Eprint {http://arxiv.org/abs/0805.0325}
  {arXiv:0805.0325 [hep-ph]} \BibitemShut {NoStop}%
\bibitem [{\citenamefont {Asztalos}\ \emph {et~al.}(2010)\citenamefont
  {Asztalos} \emph {et~al.}}]{Asztalos:2009yp}%
  \BibitemOpen
  \bibfield  {author} {\bibinfo {author} {\bibfnamefont {S.~J.}\ \bibnamefont
  {Asztalos}} \emph {et~al.} (\bibinfo {collaboration} {ADMX}),\ }\href
  {\doibase 10.1103/PhysRevLett.104.041301} {\bibfield  {journal} {\bibinfo
  {journal} {Phys. Rev. Lett.}\ }\textbf {\bibinfo {volume} {104}},\ \bibinfo
  {pages} {041301} (\bibinfo {year} {2010})},\ \Eprint
  {http://arxiv.org/abs/0910.5914} {arXiv:0910.5914 [astro-ph.CO]} \BibitemShut
  {NoStop}%
\bibitem [{\citenamefont {Du}\ \emph {et~al.}(2018)\citenamefont {Du} \emph
  {et~al.}}]{Du:2018uak}%
  \BibitemOpen
  \bibfield  {author} {\bibinfo {author} {\bibfnamefont {N.}~\bibnamefont {Du}}
  \emph {et~al.} (\bibinfo {collaboration} {ADMX}),\ }\href {\doibase
  10.1103/PhysRevLett.120.151301} {\bibfield  {journal} {\bibinfo  {journal}
  {Phys. Rev. Lett.}\ }\textbf {\bibinfo {volume} {120}},\ \bibinfo {pages}
  {151301} (\bibinfo {year} {2018})},\ \Eprint
  {http://arxiv.org/abs/1804.05750} {arXiv:1804.05750 [hep-ex]} \BibitemShut
  {NoStop}%
\end{thebibliography}%

\end{document}